\documentclass[prd,aps,floats,amssymb,preprint]{revtex4}

\newcommand{\beq}[1] {\begin{equation}\label{#1} }
\newcommand{\eeq} {\end{equation} }

\newcommand{\bea}[1]{\begin{eqnarray}\label{#1} }
\newcommand{\eea}{\end{eqnarray}}

\newcommand{\al}{\alpha}

\newcommand{\sig}{\sigma}

\def\DESepsf(#1 width #2){\epsfxsize=#2 \epsfbox{#1}}

\begin{document}

\title{Relaxing Cosmological Constraints on Large Extra Dimensions}
\author{Cosmin Macesanu\footnote{cmacesan@physics.syr.edu} and 
Mark Trodden\footnote{trodden@physics.syr.edu}}
\affiliation{Department of Physics \\
Syracuse University \\
Syracuse, NY 13244-1130, USA}

\begin{abstract}
We reconsider cosmological constraints on extra dimension theories from the excess 
production of Kaluza-Klein gravitons. We point out that, if the normalcy temperature is above $1$ GeV, then 
graviton states produced at this temperature will decay early enough that
they do not affect the present day dark matter density, or the diffuse 
gamma ray background.
We rederive the relevant cosmological constraints for this scenario.
\end{abstract}

\vspace*{-0.5cm}
\begin{flushright}
SU-4252-798
\end{flushright}
\vspace{0.5cm}

\maketitle

\section{Introduction}
\label{sec:intro}
Beyond the three spatial dimensions we observe may lie many others, with total volume small 
enough to have escaped detection through microphysical or cosmological measurements. While
such a proposal is not new~\cite{Kaluza:tu,Klein:tv}, there exist a host of contemporary incarnations~\cite{Antoniadis:1990ew,Lykken:1996fj,Arkani-Hamed:1998rs,Antoniadis:1998ig,Randall:1999vf,Lykken:1999nb,Arkani-Hamed:1999hk,Antoniadis:1993jp,Dienes:1998vg,Kaloper:2000jb} that allow the extra dimensions to have a significantly larger spatial extent than had previously
been imagined. The central feature in these new constructions is the idea that standard model
particles may be confined to a $3+1$ dimensional submanifold - or brane - while gravitational
degrees of freedom may propagate in the entire space - the bulk. Such an approach liberates the
Kaluza-Klein (KK) idea from the strong constraints posed by precision laboratory and collider 
measurements of the electroweak theory and opens up new avenues for addressing 
long-standing particle physics and cosmological problems.

In place of traditional constraints, large extra dimension models face a set of new issues at the
high energy frontier, both through collider experiments and cosmology. A particularly general class of cosmological constraints arise from the overproduction of
Kaluza-Klein gravitons. In the four dimensional effective theory describing our universe for most
of its history, the matter content consists of fields confined to the brane, the graviton zero mode
(playing the role of our graviton) and a tower of massive graviton excitations having non-zero
momenta in the extra dimensions. Cosmological constraints arise because standard model particles
at high energy may create these KK gravitons. Since these particles are massive and long-lived (having
only gravitational strength couplings), their overproduction can lead to them dominating the universe
and coming into conflict with observations. This can occur indirectly, for example they may interfere with 
primordial nucleosynthesis, or directly, for example they may overclose the universe.

In this paper we reconsider cosmological constraints arising from the overproduction of KK
gravitons. Previous studies~\cite{Hall:1999mk,Hannestad:2001nq,Fairbairn:2001ab,Fairbairn:2001ct} concentrate
on scenarios in which  the temperature at which graviton production effectively
starts (the so-called normalcy temperature $T_*$) is in the
the MeV range. Then the produced gravitons are long-lived, and the most 
stringent cosmological limits came from constraints  on the present day
dark matter density and on the diffuse gamma ray background.
We note that when the normalcy
temperature is in the GeV range or above, the gravitons produced at these
temperatures decay before recombination time. 
We then find that the above constraints are significantly ameliorated
by such decays. Note also that astrophysical 
constraints~\cite{Cullen:1999hc,Barger:1999jf,Hannestad:2001xi,Hannestad:2001jv}
will be relaxed too, since in our scenario the fundamental scale of gravity
is larger then in previous analyses (For another way to avoid these constraints see~\cite{Mohapatra:2003ah}). However, new constraints become 
significant, the most stringent one coming from the 
requirement that the graviton decay products do not destroy the predictions
of the abundances of light elements 
created during Big-Bang Nucleosynthesis (BBN). 

The structure of this paper is as follows. In the next section we briefly describe the main theoretical 
framework of the model. In section~\ref{density} we (re)compute 
the standard cosmological
constraints on KK graviton production for the case when the
graviton decays are negligible. Then, in~\ref{decays} we introduce the effects of decays and rederive
the relevant constraints and in section~\ref{otherconsts} we briefly address the effect of early KK graviton production on the evolution of other
cosmological parameters. In section~\ref{softening} we consider the 
possibility that the graviton-matter interaction at high energies
might be modified, for example in soft or fat brane scenarios, and in section~\ref{conclusions} we offer a summary and our concluding comments.

\section{The Model}
\label{sec:model}
The general framework consists of a $4+d$ dimensional spacetime, with $3+1$ dimensions 
corresponding to those we are familiar with, and the extra $d$ spatial dimensions compactified. 
We follow the conventions that $M,N,\ldots = 0,1,\ldots ,4+d$,
$\mu,\nu,\ldots = 0,1,2,3$ and $a,b,\ldots = 5,6,\ldots ,4+d$.
For simplicity, in this paper we shall assume compactification on a d-torus of common radius $r/2\pi$, 
although other geometries may also be
studied~\cite{Kaloper:2000jb,Starkman:2000dy,Starkman:2001xu,Nasri:2002rx}. Writing the
bulk metric as $G_{MN}$, we define the linearized metric
$H_{MN}$, describing the gravitational degrees of freedom, by $G_{MN}=\eta_{MN}+H_{MN}$.
Gravity propagates in the entire bulk and so it is convenient to expand the linearized metric as
\begin{equation}
\label{gravitonmodes}
H_{MN}(x,y)=\sum_{\overrightarrow{n}} H_{MN}^{\overrightarrow{n}}(x) 
\exp\left(i\frac{2\pi \overrightarrow{n}\cdot\overrightarrow{y}}{r}\right) \ ,
\end{equation}
where $x^{\mu}$ are brane coordinates and $y^a$ are those in the bulk.

This field may be decomposed into its scalar, vector and tensor components with respect to the $3+1$ 
dimensional Poincar\'{e} group as
\begin{equation}
\label{4ddecomposition}
H_{MN}=\frac{1}{\sqrt{V_d}}\left(
\begin{array}{cc}
h_{\mu\nu}+\eta_{\mu\nu}\phi      &   A_{\mu a} \\
A_{\nu b}      & 2\phi_{ab} 
\end{array} \right) \ ,
\end{equation}
where $\phi \equiv \phi_a^a$ and $V_d \equiv r^d$ is the volume of the d-torus.

All other fields, and in particular those of the standard model, are restricted to propagate only on 
the brane, so that they do not have equivalent Kaluza-Klein excitations.

\section{Density of KK Gravitons}
\label{density}
As mentioned above, we are interested in the possibility that KK gravitons may be produced through
high-energy processes involving standard model particles on our 3-brane. If such processes are 
abundant at high cosmic temperatures, then the cooling of the universe will be radically different from the
usual effect of cosmic expansion.

The Kaluza-Klein gravitons are the tower of four dimensional 
excitations, described by~(\ref{gravitonmodes}), of the field $h_{\mu\nu}$, defined in~(\ref{4ddecomposition}) (the vector gravitons $A_{\mu a}$ do not
couple with Standard Model matter, and we neglect 
production of scalar gravitons $\phi_{ab}$). The Feynman rules for the
matter-graviton interaction have been derived in 
\cite{Han:1998sg,Giudice:1998ck}.
 Our goal is to compute the density of these particles produced at relevant
epochs during the expansion of the universe.

To simplify our analysis, let us consider a particular graviton state  of mass 
$m = 4 \pi^2 {\overrightarrow{n}}^2/r^2$, and label the state
by its mass, rather than by its KK vector. The number density of these gravitons evolves according
to
\beq{e1} 
\dot{n}_m + 3n_m H = P_m - \Gamma_m n_m  \ ,
\eeq
where $P_m$ is the production rate and $\Gamma_m$ is the decay rate.  In order to compute the 
density of gravitons of mass $m$ produced in a given period of time, one must integrate Eq. (\ref{e1}).
To achieve this it will be convenient, as usual, to transform the equation in two ways. First, we 
change the dependent variable from cosmic time to temperature, using the time-temperature 
relation which holds during the radiation dominated era 
\begin{equation}
t = {1.5 \over \sqrt{g_*}} \ { \bar{M}_{p} \over T^2} \ ,
\end{equation}
(since generally only gravitons produced during this epoch are relevant).
Here $ \bar{M}_{p} \equiv (8\pi G)^{-1/2}$ is the reduced Planck mass. 
Second, we introduce the scaled number density $Y_m \equiv n_m/T^3$. Denoting by a prime
differentiation with respect to temperature, equation~(\ref{e1})
now becomes
\begin{equation}
\label{ymdiffeq}
{Y'}_m =\frac{3}{\sqrt{g_*}}\frac{\bar{M}_p}{T^3}\left(\Gamma_m Y_m - \frac{P_m}{T^3}\right) \ .
\end{equation}

Now, there are two types of processes that contribute to the right hand side of this equation. 
The first is inverse decay, occurring when the graviton is generated, for example, 
via neutrino-antineutrino annihilation or 
through photon-photon interactions
\begin{eqnarray}
\nu \bar{\nu} \rightarrow G_m & \ \ \ & \gamma \gamma \rightarrow 
G_m \ .
\end{eqnarray}
The second type of interaction is graviton radiation, 
generated for example through
\beq{aa}
 e^+ e^- \rightarrow \gamma G_m \ \ \ \   e^- \gamma \rightarrow e^- G_m
\eeq
Note that these are just illustrations of the types of processes, not a
full enumeration, and that the actual processes depend on the temperature $T$ at which
the production takes place.
 
Let us begin with inverse decays and, for simplicity, ignore the subsequent decays of the gravitons 
produced in this way. Recalling that the spin-summed amplitude squared for
$\nu \bar{\nu} \rightarrow G_m$ is  $s^2/4 \bar{M}_{p}^2$, 
the 
production rate for this process is given by
\begin{equation}
\label{invprodrate}
P_m(\nu \bar{\nu} \rightarrow G_m) =  {m^5 T \over 128 \pi^3  \bar{M}_{p}^2}
{\cal K}_1 \left({m \over T}\right) \ .
\end{equation}
 Here
${\cal K}_1$ is the modified Bessel function of the second kind, with
asymptotic behavior
\begin{equation}
{\cal K}_1(z) \sim \left\{ 
\begin{array}{ll}
z^{-1} & \ \hbox{for}\  z \ll 1 \\
\sqrt{\pi \over 2 z} \ e^{-z} & \ \hbox{for}\  z \gg 1 
\end{array} \right. \ .
\end{equation}
As one would expect, the exponential suppression for $m > T$ implies that is not possible
to produce gravitons whose mass is much larger than the temperature.

Denoting the temperature at which production commences by $T_i$, the 
number density of gravitons at a lower temperature $T_f$ is obtained by integrating~(\ref{e1}),
using~(\ref{invprodrate}), yielding
\beq{ey}
\label{invprodintegral}
Y_m(T_f) \simeq {10^{-3}\over \sqrt{g_*}} {m\over  \bar{M}_{p} }
\int_{m/T_i}^{m/T_f} dz\ z^3 {\cal K}_1(z) \ .
\eeq

We may perform a similar calculation for graviton radiation processes, for which
the graviton production rate is
\begin{equation}
P_m (a\ b \rightarrow c\ G_n)  = \langle \sig v \rangle n_a n_b \ ,
\end{equation}
where $n_a, n_b $ are the number densities of particles in the
initial states at temperature $T$
($n_{\gamma} \simeq 2.4\ T^3/\pi^2, n_f \simeq 1.8\ T^3/\pi^2$ for
relativistic bosons and fermions respectively).
 The thermally averaged cross-section is 
\beq{av_sig}
\langle \sig v \rangle  \ =  \
{T^6 \over 16 \pi^4  n_a n_b } \int_{m/T}^{\infty} dz \ z^4 {\cal{K}}_1(z)
\sigma(z^2 T^2) \ ,
\eeq
with $ z T = \sqrt{s}$ the center of mass (CM) energy at which the collision takes place.
Taking  $ \langle \sig v \rangle \simeq \al / \bar{M}_{p}^2$ 
as a general approximation valid for these types of processes,
it is then straightforward to integrate~(\ref{e1}) to obtain
\beq{yr} 
Y_m(T_f) \simeq {12 \al \over \pi^4 \sqrt{g_*}} { T_i - T_f\over \bar{M}_{p} } 
\simeq {12\al \over \pi^4 \sqrt{g_*}} { T_i\over \bar{M}_{p} } \ ,
\eeq
where the final step merely acknowledges that, generally, $T_f \ll T_i$. It is worth commenting
that if $m > T_f$, then $T_f$ should be replaced with $m$ 
in the above expression, while if $m \gg T_i$, the result will be 
close to zero, 
due to the exponential suppresion in (\ref{av_sig}).
Also note that if the mass of the graviton is of the same order of magnitude 
as $T_i$, then the graviton densities (\ref{ey}), (\ref{yr})   generated by 
the two types of
processes (inverse decay and radiation) 
are roughly of the same order magnitude, while if $T_i \gg m$  the
graviton radiation type will dominate.

In order to compute the total graviton density, it remains to 
sum over all the KK excitations of the graviton via
\begin{equation}
\rho_G = \sum_{\overrightarrow{n}}  m_{\overrightarrow{n}} [T_0^3 Y_{\overrightarrow{n}}(T_0)] \ .
\end{equation}
To accomplish this we replace
the sum by an integral via
\begin{equation}
\sum_n \rightarrow  S_{d} { \bar{M}_{p}^2 \over M_D^{2+d}} 
\int_0^{m_{\rm max}} m^{d-1} dm \ ,
\end{equation}
where $M_D$ is the fundamental Planck scale in the full $4+d$ dimensional theory,
defined through $ \bar{M}_p^2 = M_D^{d+2} (r/2\pi)^d$, and $S_d = 2 \pi^{d/2}
/\Gamma (d/2)$ is the surface area of the
unit sphere in $d$ dimensions.

Since we wish to compute the present day graviton energy density, we 
take $T_f = T_0 \sim 2 K$ in (\ref{ey}), yielding
\begin{equation}
\label{roint}
\rho_G = S_{d} T_0^3 { \bar{M}_{p}^2 \over M_D^{2+d}} \ {f\over \bar{M}_{p} }
\int_0^{m_{\rm max}} dm\ m^{d+1} \int_{m/T_i}^{\infty} dz\ z^3 {\cal K}_1(z) \ ,
\end{equation}
where $f = 10^{-3}/\sqrt{g_*} \simeq 3\times 10^{-4}$ (taking 
$g_* \simeq 10$, valid for temperatures smaller than 1MeV) and where, since 
$m \gg T_0$,we have approximated the upper limit of the integral in~(\ref{invprodintegral}) 
by $\infty$. 

Usually one would also take $m_{\rm max} \rightarrow \infty$, 
since the contribution from higher mass states will be supressed
by the exponential decay of the integral over ${\cal K}_1$.
However, because of this effective cutoff, we shall instead introduce an effective maximal mass 
$m_{\rm max}\equiv r T_i$, with $r$ a phenomenological constant, and approximate the lower
limit of the integral in~(\ref{roint}) by zero so that the double integral in~(\ref{roint}) becomes
\begin{equation}
\int_0^{r T_i} dm\ m^{d+1} \int_{0}^{\infty} dz\ z^3 {\cal K}_1(z) 
=\frac{3\pi}{2} {(r T_i)^{d+2} \over d+2} \ .
\end{equation}
It is easily checked that $r\sim 6 $ 
yields a good fit to the exact results obtained
by numerical integration (for $d=6$ $r$ will be slightly higher
than 6, while for $d = 2$ slightly lower). 
Note that the highest mass
which can be produced effectively is somewhat larger that what one might
naively assume, namely $2T_i$. This is to be expected, since there exists 
a large polynomial enhancement in the number of KK states available, 
which partially compensates for the Boltzmann suppression. 

We then obtain 
\beq{rores}
 \rho_G \simeq {3 \pi \over 2} f {S_{d}\over d+2} \bar{M}_{p} T_0^3 
\left({ r T_* \over M_D} \right)^{d+2} \  ,
\eeq
where we have denoted the temperature at which effective graviton production 
starts by $T_*$ (also called normalcy temperature).
Requiring that the fraction of the cosmological critical density in KK gravitons $\Omega_G$ be less than 
that in matter ($\Omega_G < 0.3 $) then implies
\beq{rhor}
5\times 10^{-5}\ {3 \pi \over 2} f {S_{d}\over d+2} { \bar{M}_{p} \over T_0} 
\left({ r T_* \over M_D} \right)^{d+2} < 0.3 
\eeq
or 
\begin{equation}
\left({ r T_* \over M_D} \right)^{d+2} < 0.5 \times 10^7 {T_0\over \bar{M}_{p} } 
\sim 0.5 \times 10^{-24} \ 
\end{equation}
(here we have taken $S_{d}/ (d+2) \sim 1 $ for all $d$).
Thus, in the case of $d=2$, we obtain 
$ r T_*/M_D < 10^{-6}$, so that if $T_* = 1$ MeV, then $M_D > 6$ TeV. 

Note that
in Eq. (\ref{rhor}) we have taken into account the contribution 
of only one type of neutrino. It is necessary to multiply the left hand side of
the equation by a factor $R_c$, which takes into account the number of channels
through which this process can proceed and the relative strengths of the cross
sections in these channels. For example, if the gravitons are produced 
at energies lower than 1MeV, then $R_c = 7$, where a factor of 3 comes from 
the three neutrino families, and an additional factor of 4 arises because 
the $\gamma \gamma \rightarrow G_m$ annihilation 
cross-section is 4 times larger than the neutrino one.

Now let us move on to evaluate the graviton density created by radiation type processes.
Using (\ref{yr}), we obtain
\begin{equation}\label{roint3}
 \rho_G  = 
S_{d} T_0^3 { \bar{M}_{p}^2 \over M_D^{2+d}} \ {{\tilde f}  \over \bar{M}_{p} } 
\int_0^{m_{\rm max}} dm\ T_* m^{d} \ ,
\end{equation}
where ${\tilde f} \equiv {12\al / (\pi^4 \sqrt{g_*})} \simeq  4 \times 10^{-4}$.
 Choosing 
the same upper limit of integration, $m_{\rm max} = r T_*$, as for the previous case
yields 
\begin{equation}
\label{roG2}
\rho_G \simeq  {{\tilde f} \over r} {S_{d}\over d+1} \bar{M}_{p} T_0^3 
\left({ r T_* \over M_D} \right)^{d+2} \ ,
\end{equation}
which is an order of magnitude smaller than the contribution coming from
$\nu \bar{\nu}$ annihilation. However,  this conclusion holds
for gravitons produced at low temperatures. If $T_* > 300 $ MeV,
then gravitons can also be produced, for example, through $q \bar{q} \rightarrow g G_m,
q g \rightarrow q G_m $ and $ g g \rightarrow g G_m$ processes, for which
the cross-section
is proportional to the strong coupling constant rather than the 
electroweak one, and these contributions may become important.
 
 \section{The Effects of Decays}
 \label{decays}
Thus far we have operated under the assumption that the gravitons produced are stable. 
We would now like to examine the validity of this approximation. 
The relevant decay process is that of gravitons into two Standard Model particles. 
To estimate this, consider the lifetime for decay into photons, given by
\cite{Han:1998sg}
\begin{equation}
\tau_{\gamma \gamma} \ = \ 3 \times 10^9 \ \hbox{yr} \ 
\left( {100 \hbox{MeV} \over m } \right)^3 \ .
\end{equation}
Since the age of the Universe today is about 1.5 $\times 10^{10} $yr, we are clearly
justified in neglecting decays of gravitons with mass lower than about 100 MeV.

However, higher mass gravitons will decay before the present epoch, and so
cannot contribute to the present day dark matter density.
If such gravitons decay after recombination, their decay products will contribute to
the diffuse cosmic gamma ray background. It is therefore useful to calculate
how high the mass of gravitons must be so that they decay before 
recombination at $t_{\rm rec} \simeq 5 \times 10^5$yr.

Since we are dealing with gravitons with masses in the GeV range, it is necessary to 
consider decays to gluon-gluon, lepton-lepton and quark-quark
final states involving those particles with masses at or below this magnitude. These decays obey 
$\Gamma_{g g}=\Gamma_{\gamma \gamma}$ and
$\Gamma_{f \bar{f}} = (1/2)\Gamma_{\gamma \gamma}$ for individual gluons and fermions.
Since there are one photon, 8 gluons, 5 quarks (each of 3 colors), 3 leptons and 3 neutrinos, 
the total decay width is
\begin{equation}
\Gamma_t = \left\{ 1 + 8+\frac{1}{2}\left[(3 \times 5) +3+3\right]
\right\} \Gamma_{\gamma \gamma} 
= 19.5 \Gamma_{\gamma \gamma} \ ,
\end{equation}
so that the relevant lifetime is
\begin{equation}
\tau_G \ \simeq 1.5 \times 10^5 \ \hbox{yr} \ 
\left( {1 \hbox{GeV} \over m } \right)^3 \ .
\end{equation}
Therefore, it is clear that most gravitons with mass greater than 1 GeV will decay before recombination.

We now revisit the results of the previous section in light of what we have just learned. 
To compute the ``surviving" graviton density we clearly must take an upper limit
$m_{\rm max}\simeq 1$~GeV in the integral~(\ref{roint}). Our 
result~(\ref{rores}) then becomes
\begin{equation}
\label{rores1}
\rho_G (T_{\rm rec}) \simeq \frac{3\pi}{2} f {S_{d-1}\over d+2} \bar{M}_{p} T_{rec}^3 
\left({ m_{\rm max} \over M_D} \right)^{d+2} \ .
\end{equation}

We next need to identify those processes that contribute to the production
of gravitons with mass around 1 GeV. At low temperatures, the possibilities
were $\nu \bar{\nu}, \gamma \gamma \rightarrow G_m$. At higher temperatures,
heavier particles can appear in the initial state.
 From Eq. (\ref{ey}) 
we see that, for these types of processes,  most gravitons of mass $m$ are
produced at temperatures of order $m$. 
Therefore, the initial state in
this case will also 
contain $e^+ e^-, \mu^+ \mu^-, g g$ and $q \bar{q}$ pairs, 
where $q$ stands for the three quarks with mass lower than 1 GeV ($u,d,s$).
In this case the graviton density (\ref{rores1}) should be multiplied by a 
factor $R_c = 61$ (remembering that gauge bosons contribute a factor of four
times that of neutrinos, and leptons and quarks twice as much). Also, the effective number of degrees
of freedom for graviton produced at $T \sim 1$ GeV will be $g_* = 61.75$
(assuming Standard Model particle content).

The situation will be somewhat different 
for gravitons produced in radiation type 
processes. In this case, most of the gravitons with mass $m$ will
be produced at temperatures larger than $m$ (assuming that the
normalcy temperature $T_* \gg m$). The number density~(\ref{roG2}) 
then becomes
\begin{equation}
\label{roG4}
\rho_G \simeq  {\tilde f}  {S_{d-1}\over d+1} \bar{M}_{p} T_{rec}^3 {T_* \over M_D} 
\left({ m_{\rm max} \over M_D} \right)^{d+1} \ .
\end{equation}

Strong interaction processes give the largest contribution to graviton
production through radiation. These processes are $ q \bar{q} \rightarrow g G_m $,
for which the cross-section has to be multiplied by a factor $4 N_f$ with
respect to the  $ e \bar{e} \rightarrow \gamma G_m $ cross-section
 ($N_f$ is the number
of quark flavors contributing and 4 is the color factor), 
$ q g \rightarrow q G_m $ with a factor $8 N_f$, and
$ g g \rightarrow g G_m $ with a color factor 24. If $T_* > 200$ GeV,
all quark flavors contribute, and the total multiplicative factor
for the right-hand side of Eq. (\ref{roG4}) will be $R_c = 96$, while the
number of effective degrees of freedom at production time
will then be $g_* = 106.75$.

Since $T_* \gg m_{\rm max} \simeq 1 $ GeV, most gravitons with
this mass will
be produced through radiation type processes. Thus,
requiring that $\rho_G/\rho_{\gamma} \ll 1$
at recombination implies
\beq{rhor2}   
\left({ m_{\rm max} \over M_D} \right)^{d+1} 
\ll 5 \times 10^{-28} \times {M_D \over T_*}  \ .
\eeq
If we take $T_* \sim M_D$ then,
for $d=6$, this constraint is  satisfied for
$M_D \simeq 10 $  TeV. 
In addition the present day 
graviton contribution to the dark matter density will be negligible.

A more complete calculation would entail evaluating
the contribution of the photons resulting from late graviton decays 
to the diffuse gamma ray background. 
Also, the photons resulting from graviton decays close to recombination
time ($\tau \gtrsim 10^{10}$ sec) might distort the CMB distribution from
black body spectrum \cite{silk}. 
However, we will not address this here.

\section{Other Constraints}
\label{otherconsts}
In the previous section we have seen that, as long as $M_D >10$ TeV
(for $d = 6$), 
KK gravitons may be produced at any temperature without impact
on the present day dark matter density or on the diffuse gamma ray backround. In
this section we briefly examine other possible effects 
of graviton production at temperatures
of the order of $100$ GeV or greater on  the 
cosmological parameters.

One first test is to look at the depletion of the cosmological 
photon (or other SM particle) density due to annihilation into gravitons. 
Such a depletion rate must be much smaller than the dilution due to the 
expansion of the universe
\begin{equation}
2\sum_n P_m \ll 3 n_{\gamma} H \ .
\end{equation}
This can be written explicitly as
\begin{equation} 
2 \times 10^{-3}\ S_{d-1} { T \over  M_D^{2+d}}
\int_0^{m_{\rm max}} dm \  m^{d+4} {\cal K}_1\left({m \over T}\right)
\ll  {7.2 \over \pi^2} T^3 \sqrt{{ \pi^2 \over 90} g_*}
{T^2 \over \bar{M}_{p}} \ ,
\end{equation}
which yields
\begin{equation}
\left( {r T \over M_D}\right)^{d+2} \ll 10^{-16} \left({T \over 1\ \hbox{GeV}}\right) \ .
\end{equation}
For example this gives $r T < M_D/100$ for $d = 6$. Similar constraints are obtained
if one considers the depletion of the gluon and quark densities.

 The success of the theory of primordial nucleosynthesis also
sets quite stringent constraints on the expansion rate of the universe at
$T = T_{BBN} \simeq 1$ MeV. In particular, it is necessary that the universe be
radiation dominated at that time. Therefore we require
\begin{equation}\label{bbn_ctr}
{\rho_G \over \rho_{\gamma}}(T_{BBN}) \ll  1 \ .
\end{equation}
Using~(\ref{rores}) and  (\ref{roG2}) to evaluate $\rho_G$ (and taking into account
all contributing processes) this constraint becomes
\begin{equation}\label{bbn1}
\left( {r T_* \over M_D}\right)^{d+2} \ll 5 \times 10^{-21} \ ,
\end{equation}
which is a  significantly stronger constraint than 
the previous one, yielding, for example, 
$T_* < 0.4 \times 10^{-3} M_D$ for d = 6.

It is likely, however, that the strongest constraint on such a scenario
will come from requiring that the decay products of the massive gravitons
do not destroy the light elements abundances  predicted by BBN (see, for
example~\cite{Dimopoulos:1988ue}). 
An analysis using recent data indicates that the abundance 
of a generic unstable massive particle $X$ decaying
mainly to hadrons at a time
between $10^4 \ - \ 10^{10}$ seconds has an upper limit $Y_X \lesssim 
10^{-14}/m_X$(GeV)~\cite{Kawasaki:2004yh}. This constraint is about
ten orders of magnitude more stringent than (\ref{bbn1})
\begin{equation}\label{bbn2}
\left( {r T_* \over M_D}\right)^{d+2} \lesssim  10^{-31} \ . 
\end{equation}
Constraints for the case when the heavy particle decays radiatively
(to photons) are
somewhat weaker~\cite{Cyburt:2002uv} (For some applications to specific scenarios
see~\cite{Feng:2003xh,Feng:2003uy,Feng:2003nr}). However, neither of these numbers may be
directly applicable to our case. Since the gravitons decay mostly to
hadrons, but have a sizable
decay branching ratio to electroweak gauge bosons and leptons, 
one should perform an analysis taking into account both types of decays.
This may weaken somewhat the constraint (\ref{bbn2}), since 
the overproduction of D and $^6$Li through hadronic processes (which
sets the strongest constraint on $Y_X$ in the interesting region of parameter
space) might potentially 
be compensated by a destruction of these elements due to energetic photons. 

Another potential constraint one might consider arises from the entropy production from the decays
of KK gravitons. Such production could unacceptably dilute a pre-existing baryon to entropy ratio. 
In the well-known example of gravitino decay, this
dilution factor can be as high as $10^7$ and is of real concern for most baryogenesis mechanisms. A
rough calculation in our case reveals a number that at most is of order $10^2$
(provided that (\ref{bbn1}) is satisfied). Given that a number of 
baryogenesis models are able to accommodate such a number, we shall not 
pursue this constraint further here.

Finally, we note that the constraints 
discussed in this section have the potential to 
make the observation of KK gravitons at colliders quite challenging. 
For example,
if $T_*$ is in the 10 GeV range, 
then the weaker BBN constraint (\ref{bbn1}) will
require that the fundamental gravity scale $M_D$ is in the 10 TeV range 
(for d = 6),
which still alllows for the observation of graviton effects at 
near-future colliders
(see, for example \cite{Hewett:2002hv} and references therein). 
However, if the stronger BBN constraint (\ref{bbn2}) is valid, then this
will push $M_D$ to 100 TeV range or higher.

\section{Brane Softening and Other Natural Cutoffs}

The validity of the constraints derived above assumes that the matter-graviton
interaction stays unchanged up to energies of order $T_*$. However, at energies
close to the fundamental scale of gravity new effects may appear. 

One such effect would be the softening of the gravity-matter interaction
due to brane fluctuations \cite{Murayama:2001av}.  The 
interactions derived in \cite{Han:1998sg}
assume that the SM brane is rigid. However,
this is not necessarily so; for example if the energy
is high enough that brane oscillations can be excited, then the matter-graviton
interaction vertex will aquire an effective form factor 
${\cal{F}} =  e^{ -{1\over 2} {m^2\over \Delta^2} }$, 
with the `softening scale' $\Delta $ related to the brane tension.
As a result the cross-section for production of gravitons with
mass greater than $\Delta$ will be exponentially suppressed.
 
The softening scale $\Delta$ therefore provides a natural origin for the cutoff on 
the magnitude of the masses of gravitons produced in the early universe. Assuming
that the normalcy temperature is much larger than $\Delta$, most
of the gravitons with mass of order $\Delta$ will be produced through radiation
processes, and the 
graviton energy density (\ref{roint3}) at temperature $T$  becomes
\begin{eqnarray}
\label{roint1}
 \rho_G  & = & 
S_{d} T^3 { \bar{M}_{p}^2 \over M_D^{2+d}} \ {{\tilde f}  \over \bar{M}_{p} } 
\int_0^{m_{\rm max}} dm\ T_* m^{d} e ^{-{m^2\over \Delta^2}} \ 
  \nonumber \\
& \simeq &  {\tilde f} {S_{d}\over d+1} \bar{M}_{p} T^3 
\left({ \Delta \over M_D} \right)^{d+1} \ ,
\end{eqnarray}
where we have taken $T_* \simeq M_D$. 
As we saw in the previous section, the strongest constraints
on graviton production in early universe comes from
requiring the preservation of BBN predictions. (In this
section we will use the expansion rate constraint (\ref{bbn_ctr}), since
we do not know the precise numbers for (\ref{bbn2})).
For $\Delta > 1$ GeV this constraint now becomes
\begin{equation}
\left( {\Delta \over M_D}\right)^{d+1} \ll 3. \times 10^{-21} \ .
\end{equation}
Note that there is still some hierarchy involved - the most natural scale for
$\Delta$ is close to $M_D$, while the above constraint requires several orders
of magnitude between the two quantities. However, the normalcy temperature
$T_*$ in this scenario can be as high as the fundamental gravity scale $M_D$.

An alternate possibility, often referred to as the {\it fat brane} 
scenario~\cite{DeRujula:2000he} allows the SM particles to be 
{\it localized} around a brane, rather than strictly confined to one. 
This means that they
may propagate in one or more of the extra dimensions in which gravity lives,
but only for a reduced distance $R \leq {\cal O}({\rm TeV}^{-1})$ to ensure that
the SM KK excitations satisfy collider bounds.

For simplicity assume that the confining potential is an infinite
square well, so that the wave functions of the SM particles are unity on the
brane ($ 0 < y_i <\pi R$) and zero outside. The graviton-matter vertex
function now acquires a form-factor \cite{DeRujula:2000he,Macesanu:2003jx}
\begin{equation}
{\cal{F}} = {1\over (\pi R)^d } \int_0^{\pi R} d \overrightarrow{y}
\exp\left(i\frac{2\pi \overrightarrow{n}\cdot\overrightarrow{y}}{r}\right)
\end{equation}
and the production cross-section is correspondingly multiplied by
\begin{equation}  
|{\cal{F}}|^2 = \prod_i \left( M \over m_i \pi \right)^2
4 \sin^2 \left( \pi m_i \over 2 M  \right) \ ,
\end{equation}
with $M \equiv 1/R $ and $m_i \equiv 2 \pi n_i/r$. 
The energy density of gravitons 
produced in the early universe through inverse decay processes then becomes
\begin{equation}
\label{roint2}
\rho_G(T) =  T^3 { \bar{M}_{p}^2 \over M_D^{2+d}} \ {f\over M_{p} } 
\int_{m<M_D} m^2  \prod_i dm_i\ 
\left( 2 M \over m_i \pi \right)^2
 \sin^2 \left( \pi m_i \over 2 M  \right)
\int_{m/T_i}^{\infty} dz\ z^3 {\cal K}_1(z)  \ ,
\end{equation}
where $m^2 = \sum_i m_i^2$.

Note that the result of the integral over $m$ increases linearly with the
upper limit of integration (which we take to be $M_D$). This can be seen 
by integrating each part separately
\begin{equation}
\label{int2}
\sum_i \int_{-M_D}^{M_D}  d m_i 
 \sin^2 \left( \pi m_i \over 2 M  \right)
\prod_{j \neq i} \int {d m_j \over m_j^2}
 \sin^2 \left( \pi m_j \over 2 M  \right) 
 \simeq  M_D \left( { \pi^2 \over 2 M }\right)^{d-1} d \ .
\end{equation}
Taking $T_i \gg m$ in (\ref{roint2}), we obtain
\beq{roint4}
\rho_G(T) \simeq {3 \pi  \over 2} f \bar{M}_p T^3 {d\over \pi^2}
\left( { 2M \over M_D }\right)^{d+1} \ .
\eeq

We similarly evaluate the graviton energy density created in
radiation processes to be (with $T_* \simeq M_D$):
\beq{a1}
\rho_G(T) \simeq   {\tilde{f}} \bar{M}_p T^3 { d\over \pi^2}
\left( { 2M \over M_D }\right)^{d+1} \ln \left( { M_D \over \pi M }\right) .
\eeq
The BBN expansion rate constraint (\ref{bbn_ctr})
then reads
\begin{equation}
\left( {2 M \over M_D}\right)^{d+1}  \ln \left( { M_D \over \pi M }\right)
\ll 5\times 10^{-21} \ .
\end{equation}
This again requires that the new scale $M$ (related to the
confining potential) should be several orders of magnitude smaller
than the gravity scale $M_D$, but one does not need a low normalcy 
temperature in this scenario either.

\label{softening}

\section{Conclusions}
\label{conclusions}
The possibility of large extra dimensions has opened up an entirely new set of approaches to the
problems of both particle physics and cosmology. The brane world construction is 
designed to obviate the traditional bounds on higher-dimensional versions of the 
electromagnetic and weak interactions by confining our standard model particles to the
usual $3+1$ dimensional submanifold. However, there are significant constraints on
large extra dimension models from both particle physics and cosmology.

A generic constraint arises from the potentially problematic production
of Kaluza-Klein graviton states during the evolution of the early universe. While these effects
have been considered before, in this paper we have reconsidered them in the range of normalcy
temperatures above $1$ GeV. There are good reasons to consider this a particularly attractive region of parameter space. Perhaps the most compelling candidate
for dark matter is a weakly interacting massive particle (WIMP) - a heavy particle with electroweak scale 
interaction strength.
The mass of this particle is expected to be in the 10 GeV - 10 TeV 
range (see, for example \cite{Birkedal:2004xn}) and
in order to produce a sufficient relic density of this particle as the universe cools one generally
then requires $T_* = {\cal O}({\rm GeV})$ or larger.

In this cosmologically interesting range of normalcy temperatures, we have demonstrated that one
may not neglect the effects of KK graviton decay when computing constraints from KK graviton production.
We have recomputed the relic density, taking into account the decay of these particles, and have rederived
the corresponding constraints. The results cover a larger 
region of parameter space compared to previous calculations
and open up a new window of viability for these models.

\section*{Acknowledgments}
We thank Jonathan Feng and Salah Nasri for helpful comments.
The work of CM is supported by the US Department of Energy 
grant number DE-FG02-85ER40231, and that of MT 
is supported in part by the National Science Foundation under grant
PHY-0094122 and in part by a Cottrell Scholar Award from Research Corporation.

\end{document}